\documentclass[reprint,amsmath,amssymb,aps,article]{revtex4-2}
\usepackage{graphicx}  % 插图
\usepackage{lipsum}    % 示例段落（可选）
\usepackage{dcolumn}% Align table columns on decimal point
\usepackage{bm}% bold math
\usepackage{color}
\usepackage[colorlinks,linkcolor=blue,citecolor=blue,urlcolor=blue,hyperindex,breaklinks]{hyperref}
\allowdisplaybreaks[4]
\usepackage{subfigure}
\begin{document}
\preprint{APS/123-QED}
\title{Generation of Four-Component Schrödinger Cat States via Floquet Engineering in a Hybrid Ferromagnet-Superconductor System}% Force line breaks with \\
\author{Shiwen He$^1$, Zi-Long Yang$^{1}$, Sitong Jin$^{1}$, Feng-Yang Zhang$^2$}
\email{dllgzfy@126.com} 
\author{Chong Li$^1$}
\email{lichong@dlut.edu.cn}
\affiliation{$^1$School of Physics, Dalian University of Technology, Dalian 116024, China\\
$^2$School of Physics and Materials Engineering, Dalian Minzu University, Dalian 116600, China\\
}
\date{\today}
\begin{abstract}
Four-component Schrödinger cat (4C) states are  important physical resources for 
fault-tolerant quantum computing. 
However, the generation of 4C states in solid-state platforms 
remains challenging due to stringent nonlinearity requirements. 
In this paper, a Floquet-engineering scheme is proposed for generating 4C states in a 
hybrid ferromagnet-superconductor system.
Numerical simulations show that the high-fidelity 4C states 
can be generated even if the decoherence of the  system is considered.
These results provide a scalable route to multi-component cat-state 
engineering in solid-state platforms and open new avenues for quantum computation.
\end{abstract} 

\maketitle

\section{INTRODUCTION}
Nonclassical quantum states, particularly Schrödinger cat states 
\cite{Schrödinger1935}, 
play a central role in quantum information science
\cite{Sanders_2012,YUAN20221,XIA1989281,PhysRevLett.64.2771,1993PhRvA..48.2213J,PhysRevA.51.4191,scully1999quantum}. 
They serve as a versatile platform for hardware-efficient quantum error correction 
\cite{Yu2025,Sun2014,Ofek2016}, quantum sensing 
\cite{Johnson2017,Bruno2013}, studies of 
quantum decoherence, and the quantum to classical transition 
\cite{C_P_Sun_1997,PhysRevX.13.021004,PhysRevLett.77.4887,1996Sci...272.1131M,PhysRevA.84.012121}.
While two-component cat states have been extensively explored, recent efforts have focused on
the four-component Schrödinger cat (4C) states 
\cite{Johnson2017,PhysRevResearch.5.013165,Kirchmair2013,Kikura_2025}. 
The 4C states are superpositions of four coherent states, 
$|\alpha\rangle$, $|-\alpha\rangle$, $|i\alpha\rangle$, and $|-i\alpha\rangle$ with the 
coherent amplitude $\alpha$.  
The fourfold phase-space symmetry of the 4C states enriches quantum interference patterns \cite{Lee:15},
enhances resilience against photon loss
\cite{10.1063/5.0138391,Joshi_2021,PhysRevLett.111.120501,Mirrahimi_2014},
and makes these 4C states an attractive resource for
fault-tolerant quantum computing and robust qubit encoding
\cite{https://doi.org/10.1002/andp.202300010,PhysRevA.106.042614,Hastrup:20,Heeres2017,Mirrahimi_2014,Zeng2023Jul,Zeng2025Feb}. 
Experimentally, the 4C states have been created 
for qubit storage and error protection \cite{Hofheinz2009-xj,Johnson2017,doi:10.1126/science.1243289}.
Despite these advances, generating multi-component cat states in solid-state platforms, 
particularly in ferromagnet, remains a critical challenge. 
To address this challenge, we first review well established platforms for generating two-component 
cat states, thereby offering insights into extending
these techniques to four-component systems.
\par
A variety of physical platforms have been developed for generating two-component cat states 
\cite{PhysRevA.56.4175,liu2025magnoncatstatescavitymagnonqubit,PhysRevLett.57.13,PhysRevLett.125.093603,Wang2025EffectiveAR,PhysRevA.42.1703,PhysRevA.45.5193,PhysRevLett.90.027903,PhysRevA.101.043841,PhysRevLett.126.023602,PhysRevLett.127.093602,Zhang_2021,PhysRevA.99.043837,PhysRevLett.129.063602,PhysRevA.102.011502,Qin2019,Zeng:20,LI2022,PhysRevB.57.7474,PhysRevLett.127.087203,PhysRevB.103.L100403}. 
In particular, hybrid quantum systems based on magnonics 
have important applications in the past decade due to
their potential in quantum information science, quantum technology, 
quantum sensing, nonlinear and macroscopic quantum studies
\cite{Lachance-Quirion_2019,YUAN20221,Zuo_2024,PhysRevLett.129.037205,ZARERAMESHTI20221,PhysRevA.107.023709,PhysRevA.110.013711,PhysRevA.110.053710}. 
Magnons, the quantized quasiparticles of collective spin waves in 
ferromagnetic materials such as yttrium iron garnet (YIG), have attracted considerable attention. 
Due to their high spin density, low dissipation, and long coherence times 
\cite{PhysRevLett.111.127003,PhysRevApplied.2.054002,PhysRevLett.113.156401,Uchida2010,Kajiwara2010,PhysRevLett.113.083603,Wang_2025}, 
magnons have become ideal candidates for exploring macroscopic quantum phenomena. 
Recently, indirect coupling between a 
ferromagnetic magnon and a 
superconducting qubit has been demonstrated via virtual photons of the microwave cavity 
mode \cite{doi:10.1126/science.aaa3693}. 
In the static coupling scheme, the frequencies of both the superconducting qubit and the magnon 
are far detuned from the cavity frequency. 
On the other hand, theoretical proposals indicate that a direct coupling is also possible 
\cite{PhysRevLett.129.037205}.
Therefore, the quantum states can be transferred coherently from the superconducting qubit to the magnon.
Hybrid systems integrated with magnons and superconducting qubits form a versatile platform for
supporting
single magnon control 
\cite{doi:10.1126/sciadv.1603150,jr6g-dcnp}, 
quantum information storage 
\cite{Zhang2015}, 
entangled states generation 
\cite{PhysRevLett.124.053602,PhysRevA.105.022624,hu2024steady}, 
and the magnon blockade 
\cite{PhysRevB.100.134421,PhysRevA.101.042331,PhysRevA.110.012459,liu2025dispersiveinducedmagnonblockadesuperconducting}. 
\par
Floquet engineering is the concept of tailoring a system by a periodic drive, 
and it is increasingly employed in many areas of physics. 
Floquet engineering enables controlled generation of nonclassical states and offers 
versatile means for quantum control
\cite{annurev:/content/journals/10.1146/annurev-conmatphys-031218-013423}, such as  
drive-tuned band structures
\cite{Zhou2023,Ito2023,Ghimire2011,Schubert2014,Luu2015,Vampa2015}, and nonclassical dynamics
\cite{PhysRevA.93.033853,PhysRevLett.116.163602,PhysRevA.106.012609,PhysRevLett.125.237201}.
A wide variety of novel functionalities have been enabled by its application in recent years, 
including quantum dots
\cite{PhysRevX.6.041027,PhysRevLett.121.043603}, 
cold atoms 
\cite{RevModPhys.89.011004,Eisert2015,PhysRevLett.99.220403,PhysRevLett.106.220402,PhysRevLett.119.123601,PhysRevA.100.033406}, 
Josephson junctions 
\cite{PhysRevA.99.012333,Wang_2020} 
and integrated photonics 
\cite{Zhang2019}. 
In addition to their practical applications, Floquet-driven systems have significantly advanced 
fundamental research, leading to the experimental observation of novel non-equilibrium 
phenomena, including discrete time-crystalline 
\cite{Zhang2017timecrystal,Choi2017,PhysRevLett.120.040404}
and Floquet spin-glass phases
\cite{PhysRevLett.122.143903}.
\par
In this paper, we propose a theoretical framework to generate 4C states in a 
hybrid ferromagnet-superconductor quantum system. 
Each superconducting qubit is driven by a Floquet field, with a relative phase difference 
$\phi$. 
A conditional displacement-type Hamiltonian is constructed with 
the second-order interaction terms between the magnon and each
superconducting qubit. 
Therefore, in our protocol, 
reliance on higher-order nonlinearities 
is avoided, which in turn relaxes hardware requirements. 
From the conditional displacement mechanism, effective 
dynamics enable the generation of 
multi-component cat states starting with initial product states. 
Furthermore, we derive the associated 
master equation under the single-photon loss, and the 
numerical simulations show that the engineered dynamics remain 
robust under realistic dissipation. 
The phase-space trajectory of the magnon 4C states is coherently controlled by the 
joint parity of the superconducting qubit states. 
Additionally, this phase difference sets the orientation and 
phase relations of the coherent-state constellation 
with $C_4$ symmetry emerging only at a specific phase offset.
Overall, our Floquet-engineered protocol enables 
the generation of 4C states in hybrid magnon systems, 
opening new avenues for solid‑state quantum state engineering.
\par
\section{MODEL AND HAMILTONIAN} \label{sec:2} 
\begin{figure}%[htbp] %htbp 代表图片插入位置的设置
\centering %图片居中      
\includegraphics[width=8cm]{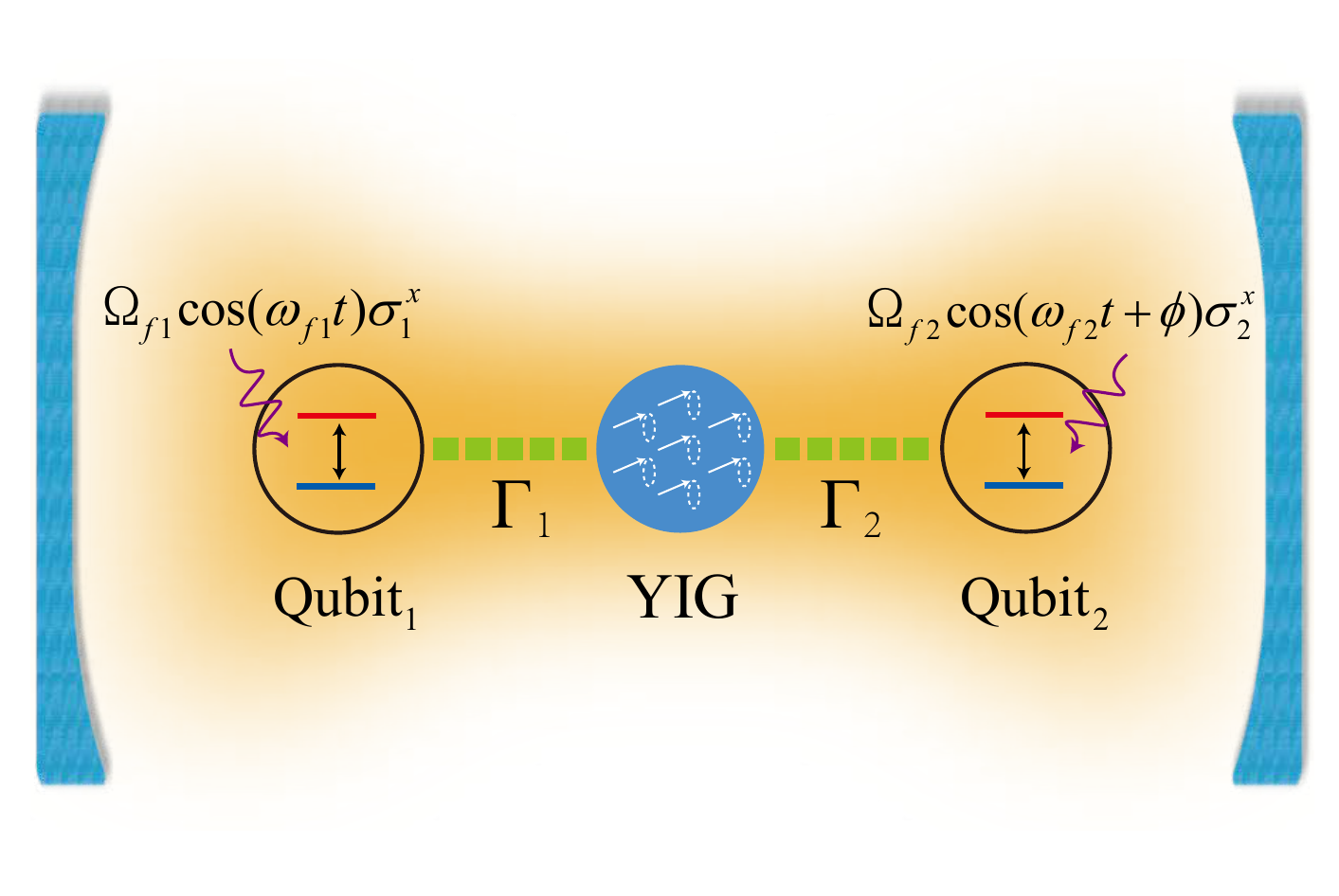} %[]中可选参数，可以设置图片的宽高      
%添加图体      
\caption{Two superconducting qubits and a YIG sphere are placed in a microwave cavity.  
Two Floquet drives ($\Omega_{f_1} \cos(\omega_{f_1} t) \sigma_1^x$, $ \Omega_{f_2} \cos(\omega_{f_2} t + \phi)
\sigma_2^x$) are applied 
to the superconducting qubits with modulation frequencies $\omega_{f_1}, \omega_{f_2}$, 
drive strengths $\Omega_{f_1}, \Omega_{f_2}$, and a tunable relative phase difference $\phi$. 
Sideband-mediated processes are controlled by the Floquet drives. 
The effective coupling strengths
$\Gamma_{1}$ and $\Gamma_{2} $ between the superconducting qubits and the magnon mode are 
induced via virtual-photon excitation of the cavity.
}
\label{fig1}
\end{figure}

We consider a hybrid quantum system composed of a YIG sphere and two superconducting 
qubits, and a microwave cavity, as illustrated in Fig. \ref{fig1}. 
Each of the two superconducting qubits is electrically coupled to the cavity mode, respectively, 
and the magnon mode is magnetically coupled to the cavity mode; 
the direct interaction between each 
superconducting qubit and the magnon mode is negligibly small \cite{doi:10.1126/science.aaa3693}. 
There is a large detuning between the magnon mode and each superconducting qubit. 
The effective coupling strengths $\Gamma_1$ ($\Gamma_2$) between the magnon mode and 
the superconducting qubit$_1$ (the superconducting qubit$_2$) are established 
by applying a Floquet drive.
The dynamics of this Floquet-engineered hybrid magnon system are governed by
the total Hamiltonian:
\begin{equation}
  \begin{aligned}
H_{\rm total} =\; & \frac{\omega_{q_1}}{2} \sigma_1^z + \frac{\omega_{q_2}}{2} \sigma_2^z + \omega_c a^\dagger a + \omega_m m^\dagger m \\
&+ g_1 \left( a \sigma_1^+ + a^\dagger \sigma_1^- \right) + g_2 \left( a \sigma_2^+ + a^\dagger \sigma_2^- \right) \\
&+ g_3 \left( m^\dagger a + m a^\dagger \right) + \Omega_{f_1} \cos(\omega_{f_1} t) \sigma_1^x \\
&+ \Omega_{f_2} \cos(\omega_{f_2} t + \phi) \sigma_2^x,
\end{aligned}
\label{equ:1}
\end{equation}
where the operators $\sigma_j^{z}$ ($j=1,2$)
are defined
by $|e\rangle_{q_j}\langle e|- |g\rangle_{q_j}\langle g|$ with the the ground state $|g\rangle_{q_j}$ and the
excited state $|e\rangle_{q_j}$ of the qubit$_j$;
$a^\dagger$ ($a$) and $m^\dagger$ ($m$) are the bosonic 
creation (annihilation) operators for the cavity and magnon modes with the 
transition frequencies $\omega_c$ and $\omega_m$, respectively; 
$\sigma_{j}^{+} = (\sigma_{j}^{x}+i\sigma_{j}^{y})/2$
and $\sigma_{j}^{-} = (\sigma_{j}^{x}-i\sigma_{j}^{y})/2$  
are the raising and lowering operators of the superconducting qubit$_j$ with frequencies
$\omega_{ q_j}$;
the terms with 
coupling strength $g_1$ ($g_2$) describe 
the interaction between the cavity mode and the superconducting qubit$_1$ (the superconducting qubit$_2$), 
while $g_3$ is the coupling strength between the magnon and the cavity mode; 
$\Omega_{f_j}$ ($\omega_{f_j}$) and $\phi$ are the Floquet driving strengths 
(Floquet driving frequencies) 
and the controllable relative phase difference. 
\par
The effect of the Floquet drives on the system dynamics 
of the system is analyzed by performing the transformation
\begin{equation}
U_1(t) = \exp\left[-i \theta_1(t) \sigma_1^x - i \theta_2(t) \sigma_2^x\right],
\label{equ:2}
\end{equation}
where $\theta_1(t) = \frac{\Omega_{f_1}}{\omega_{f_1}} \sin(\omega_{f_1} t)$ and 
$\theta_2(t) = \frac{\Omega_{f_2}}{\omega_{f_2}} \sin(\omega_{f_2} t + \phi)$ are 
defined as the dynamical phases. 
We obtain the following Hamiltonian:
\begin{equation}
\begin{aligned}
H_{\mathrm{fram}} =\; & 
\omega_c a^\dagger a + \omega_m m^\dagger m 
+ g_3 \left( m^\dagger a + m a^\dagger \right) \\
& + \sum_{j=1}^{2} \frac{\omega_{q_j}}{2} 
\left[ \sigma_j^z \cos\big(2\theta_j(t)\big) 
+ \sigma_j^y \sin\big(2\theta_j(t)\big) \right] \\
& + \sum_{j=1}^{2} \left\{ 
g_j a \left[ 
\sigma_j^+ - \frac{i}{2} \sigma_j^y 
- \frac{i}{2} \sigma_j^z \sin\big(2\theta_j(t)\big) \right. \right. \\
& \left. \left. \qquad + \frac{i}{2} \sigma_j^x \cos\big(2\theta_j(t)\big) 
\right] + \mathrm{h.c.} \right\},
\end{aligned}
\label{equ:3}
\end{equation}
where $\mu_j = 2 \Omega_{f_j}/\omega_{f_j}$ denotes a dimensionless Floquet parameter.
\par
The time-dependent coefficients in Eq. (\ref{equ:3}) are expanded
via the Jacobi-Anger expansion with Bessel functions terms. 
This expansion permits a sideband-resolved analysis of the dynamics.
Under the rotating-wave approximation (RWA), by selecting a specific Floquet drive 
modulation frequency,
the frequency of a selected sideband is set much lower than the others, so that
the effective dynamics is dominated by the $2n_0-1$-order sideband. 
We obtain the
approximate Hamiltonian 
\begin{equation}
\begin{aligned}
H^{\prime}_{\rm fram}  = & -G_1\sigma_1^z \left( a e^{-i \delta t} + a^\dagger e^{i \delta t} \right) \\
& -  G_2\sigma_2^z \left(
a^\dagger e^{i \delta t} e^{-i\Phi}
+ a e^{-i \delta t} e^{i\Phi} \right)\\
& + g_3 \left( m a^\dagger e^{i\Delta_{\rm cm }t} + m^\dagger a e^{-i\Delta_{\rm cm }t} \right),
\end{aligned}
\label{equ:4}
\end{equation}
where $G_1= g_1 J_{2n_0 - 1} \left( \mu_1 \right)/2$ and
$G_2 = g_2 J_{2n_0 - 1} \left( \mu_2 \right)/2$ 
represent the effective (Floquet-renormalized) qubit-cavity coupling strengths; 
and $\Phi = (2n_0 - 1)\phi$ corresponds to the effective relative phase difference. 
The detailed derivation can be found in the Appendix \hyperref[appendix:A]{A}.
\par
We reformulate the Hamiltonian $H^{\prime}_{\rm fram}$
as $H^{\prime}_{\rm fram} = U_{\rm aux}^{\dagger}(t) \tilde{H}_{\rm fram} U_{\rm aux}(t) - 
i U_{\rm aux}^{\dagger}(t) \dot{U}_{\rm aux}(t)$, where 
$U_{\rm aux}(t) = \exp(-i H_0 t)$ is a unitary operator 
with the auxiliary free Hamiltonian $H_0 = \delta a^\dagger a + (\delta - \Delta_{\rm cm}) m^\dagger m$;
and $\tilde{H}_{\rm fram}$ 
denotes a time-independent Hamiltonian in the auxiliary picture. 
Moreover, 
under the condition $\| V\| \ll\| H_0\| $, 
such a reformulation enables a systematic perturbative analysis.
The Hamiltonian $\tilde{H}_{\rm fram}$ is decomposed as
$ H_0 + V $ with the interaction part
$V = -G \sigma_1^z (a + a^\dagger) - G \sigma_2^z \left( a^\dagger 
e^{-i \Phi} + a e^{i \Phi} \right) + g_3 (m a^\dagger + m^\dagger a)$, 
where we have set 
$G_{1} = G_{2} \equiv G $.
\begin{figure}%[htbp] %htbp 代表图片插入位置的设置
\centering %图片居中      
\includegraphics[width=7.8cm]{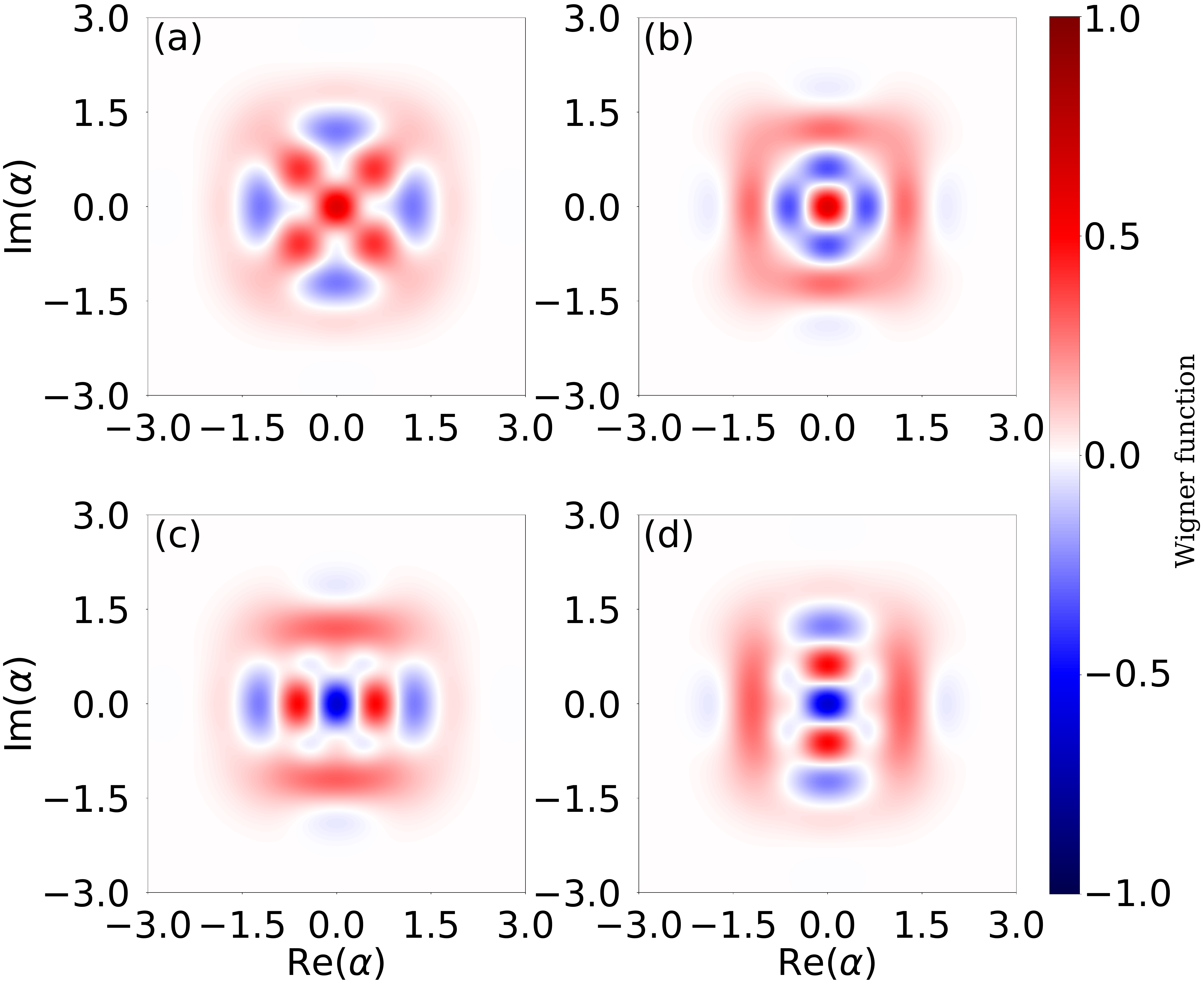} %[]中可选参数，可以设置图片的宽高      
%添加图体      
\caption{
Wigner functions of the 4C magnon states 
(a) $|\psi_{pp}\rangle_m$ 
(b) $|\psi_{mm}\rangle_m$ 
(c) $|\psi_{pm}\rangle_m$ 
(d) $|\psi_{mp}\rangle_m$ 
at $t = 40$ ns with the coherent amplitude $|\alpha| = 1.007$.
The parameters are set as: $\omega_f/2\pi = 5.0023\mathrm{GHz}$, 
$\Omega_f/\omega_f = 0.92$, $\omega_c/2\pi = 4.827\mathrm{GHz}$, 
$\omega_m/2\pi = 5\mathrm{GHz}$, $\omega_{q_1}=\omega_{q_2} = 0$, 
$g_1/2\pi = g_2/2\pi = 120\mathrm{MHz}$, $g_3/2\pi = 20\mathrm{MHz}$, $n_0=1$. 
\label{fig2}
}
\end{figure}
Therefore, 
by performing the Nakajima transformation, 
we can derive a reduced effective Hamiltonian.
The effective Hamiltonian $\tilde{H}_{\rm fram}$ can be made diagonal to 
first order in perturbation $V$ by choosing 
the anti-Hermitian generator $S$ as 
\begin{equation}
\begin{aligned}
S =& -\frac{G}{\delta} \sigma_1^z (a^\dagger - a) - \frac{G}{\delta} \sigma_2^z 
\left( a^\dagger e^{-i\Phi} - a e^{i\Phi} \right) \\
&+ \frac{g_3}{\Delta_{\rm cm}} \left( m a^\dagger - m^\dagger a \right),
\end{aligned}
\label{equ:5}
\end{equation}
such that $V+[S,H_0]=0$. Therefore, we have 
\begin{equation}
\begin{aligned}
H_{\text{eff}} =& H_0 + \frac{1}{2}[S, V] \\
=& \xi m^\dagger m  +\Gamma_3 \sigma_1^z \sigma_2^z 
+ \Gamma_1 \sigma_1^z (m + m^\dagger) \\
& +  \sigma_2^z \left( \Gamma_2 m  + \Gamma_2^{*} m^\dagger  \right),
\end{aligned}
\label{equ:6}
\end{equation}
where $\Gamma_1 = \frac{1}{2} \left( g_3 G/\Delta_{\rm cm}+ g_3 G/\delta \right)$
($\Gamma_2 = \Gamma_1 e^{i\Phi}$) 
are the effective conditional displacement-type coupling strength between 
the superconducting qubit$_1$ (the superconducting qubit$_2$) and the magnon mode; 
$\Gamma_3 = - 2G^2\cos(\Phi)/\delta $ is the effective two-body Ising interaction strength between 
the two superconducting qubits; 
$\xi =  \delta - \Delta_{\rm cm} - g^2_3/\Delta_{\rm cm}$ 
is the effective magnon transition frequency in the auxiliary frame, including the Lamb shift 
correction $g_3^2m^{\dagger}m/\Delta_{\rm cm}$.
\par
Obviously, in Eq. (\ref{equ:6}), the cavity mode is decoupled. 
The role of the microwave cavity is not limited to mediating virtual-photon 
excitations. Within the Floquet framework, higher-order sidebands are far off-resonant 
and are neglected under the RWA. When the cavity mode is adiabatically eliminated, these highly 
off-resonant contributions are further suppressed, ensuring that the effective Hamiltonian is 
dominated by the desired interaction processes. This mechanism highlights the dual role of the cavity, 
both as a mediator of virtual couplings and as a natural filter that suppresses unwanted higher-order 
sidebands.
\par
In the rotating frame $U_{\rm int}(t) = {\rm exp}({-i\xi m^{\dagger}m t})$, 
we can obtain
\begin{equation}
\begin{aligned}
\tilde{H}_{\rm eff} = \Gamma_1\left(mAe^{-i\xi t} + m^{\dagger}A^{\dagger}e^{i\xi t}\right)
+\Gamma_3\sigma_{1}^{z}\sigma_{2}^{z},
\end{aligned}
\label{equ:7}
\end{equation}
where the joint qubit-control operator $A = \sigma_1^z + \sigma_2^z e^{i\Phi}$ 
is introduced to combine the influence of 
the two superconducting qubits on the magnon 
displacement in a single operator.
\section{dynamics GENERATION OF THE four-component CAT STATES}
The time evolution under $\tilde{H}_{\rm eff}$ is obtained via a Magnus expansion \cite{PhysRevA.93.033853}, 
yielding the unitary operator:
\begin{equation}
\begin{aligned}
\tilde{U}(t) =& \exp[i\Theta(t)]\exp [\eta_1(t)m^{\dagger}A^{\dagger}-\eta_1^{*}(t)mA \\
& + i\eta_{2}(t)\sigma_1^{z}\sigma_2^{z}]
\end{aligned}
\label{equ:8}
\end{equation}
where $\Theta(t) = 2(\Gamma_1/\xi)^2[\xi t - \sin(\xi t)]$ is a global phase factor; 
$\eta_1(t) = (\Gamma_1/\xi)(1 - e^{-i\xi t })$ denotes the displacement amplitude of the magnon; and 
$\eta_2(t) = 2(\Gamma_1/\xi)^{2}\cos(\Phi)[\xi t - \sin(\xi t)]-\Gamma_3 t$ 
represents a predictable local phase that depends on the Floquet drive phase difference 
$\phi$ and the two-qubit parity operator $\sigma_1^{z}\sigma_2^{z}$.
When $\Phi = (2k+1)\pi/2$ ($k\in \mathbb{Z} $), the phase 
factor $\eta_2(t)$ is suppressed, and the two-body Ising interaction term 
$\Gamma_3 \sigma_1^z \sigma_2^z$ is eliminated, resulting in pure conditional displacement mechanism. 
In the following, we choose $k = 0$. 
\par
We consider that the initial state of
the system is $|\psi(0)\rangle = |+\rangle_{q_1}|+\rangle_{q_2}|0\rangle_m$, 
where $|+\rangle_{q_j} = (|g\rangle_{q_j} + |e\rangle_{q_j})/\sqrt{2}$ 
($|-\rangle_{q_j} = (|g\rangle_{q_j} - |e\rangle_{q_j})/\sqrt{2}$) is the eigenstate
of $\sigma_{x_j}$ with eigenvalue $+1$ ($-1$); 
and $|0\rangle_m$ is the vacuum 
state of the 
magnon mode. 
\begin{figure*}[t] % 使用 figure* 跨栏图
  \centering
  \includegraphics[width=0.9\textwidth]{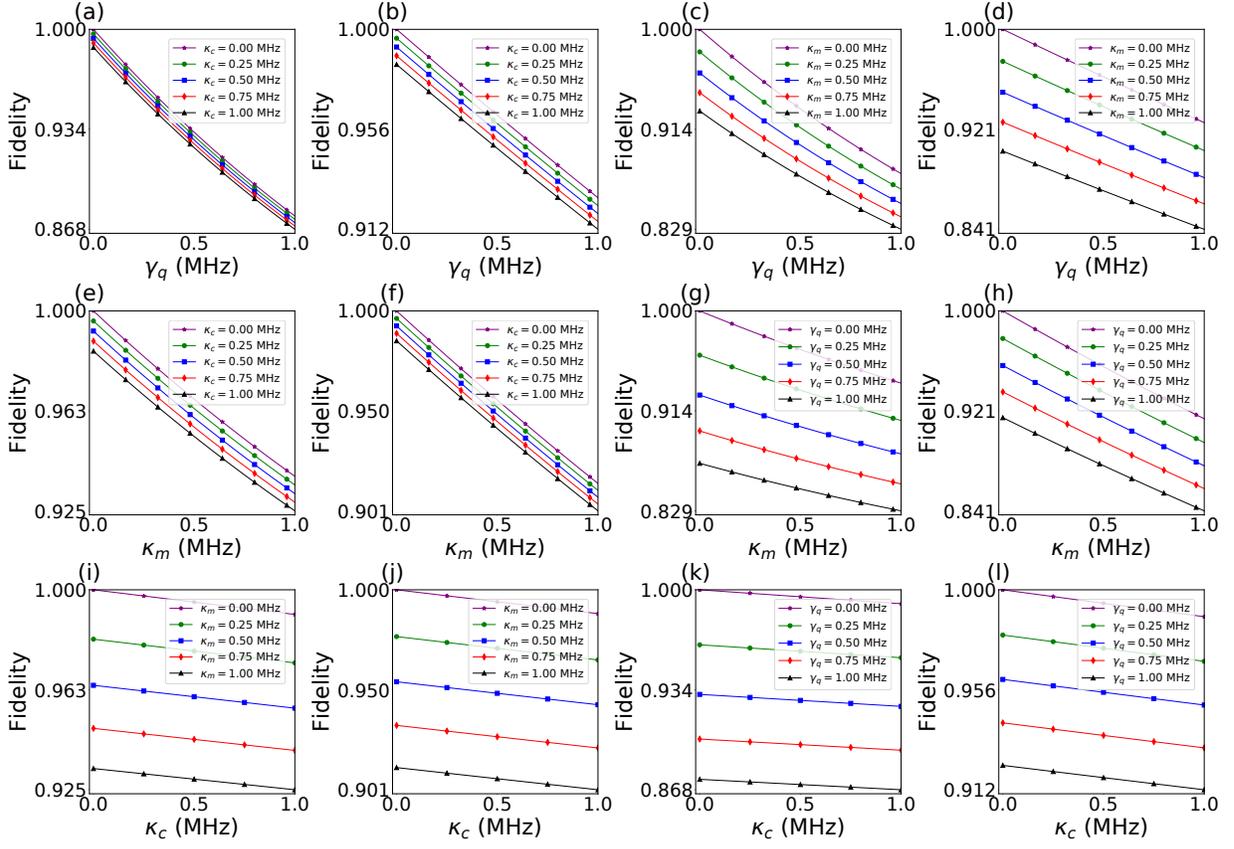} % 设为接近整页宽
  \caption{
  Fidelity of the 4C states under different dissipation conditions. 
  (a), (c), (e), (g), (i), and (k) show the fidelity of the state $|\psi_{pp}\rangle_m$, 
  while (b), (d), (f), (h), (j), and (l) correspond to the state $|\psi_{mm}\rangle_m$. 
  Figs. (a)-(d) show the dependence on the qubit dissipation rate $\gamma_q$,
  Figs. (e)-(h) show the dependence on the magnon dissipation rate $\kappa_m$,
  Figs. (i)-(l) show the dependence on the cavity dissipation rate $\kappa_m$.
  In each subplot, the remaining dissipation channel is fixed at $0.5$ MHz, 
  and different colored curves 
  in each subplot represent varying values of the fixed dissipation rate (0 MHz, 0.25 MHz, 0.5 MHz, 0.75
  MHz, and 1.0 MHz). 
  Other parameters are the same as those in Fig. \ref{fig2}.
  }
  \label{fig3}
\end{figure*}
The system state at time $t$, 
evolved via the unitary evolution operator
$\tilde{U}(t)$, 
is given by 
\begin{equation}
\begin{aligned}
|\psi(t)\rangle =&|+\rangle_{q_1} |+\rangle_{q_2} |\psi_{pp}\rangle_m +
|+\rangle_{q_1} |-\rangle_{q_2} |\psi_{pm}\rangle_m \\ 
&+|-\rangle_{q_1} |+\rangle_{q_2} |\psi_{mp}\rangle_m 
+ |-\rangle_{q_1} |-\rangle_{q_2} |\psi_{mm}\rangle_m,
\end{aligned}
\label{equ:9}
\end{equation}
where the 4C states are defined as
\begin{equation}
\begin{aligned}
|\psi_{pp}\rangle_m =&\mathcal{N}_{pp}(|\alpha(t)\rangle_m + |i\alpha(t)\rangle_m 
+|-i\alpha(t)\rangle_m  \\ &+ |-\alpha(t)\rangle_m), \\
|\psi_{pm}\rangle_m =&\mathcal{N}_{pm}(|\alpha(t)\rangle_m - |i\alpha(t)\rangle_m 
+|-i\alpha(t)\rangle_m \\ & - |-\alpha(t)\rangle_m), \\
|\psi_{mp}\rangle_m =&\mathcal{N}_{mp}(|\alpha(t)\rangle_m + |i\alpha(t)\rangle_m 
-|-i\alpha(t)\rangle_m  \\ &- |-\alpha(t)\rangle_m), \\
|\psi_{mm}\rangle_m =&\mathcal{N}_{mm}(|\alpha(t)\rangle_m - |i\alpha(t)\rangle_m 
-|-i\alpha(t)\rangle_m  \\ &+ |-\alpha(t)\rangle_m), \\
\end{aligned}
\label{equ:10}
\end{equation}
with the coherent amplitude $\alpha(t) = (1 - i)\eta_1(t)$. 
Here we have the 
normalization constants 
\begin{equation}
\begin{aligned}
&\mathcal{N}_{pp}= [4 + 4 e^{-2|\alpha|^2} + 8 e^{-2|\alpha|^2}\cos(|\alpha|^{2})]^{-1/2},\\
&\mathcal{N}_{pm} = \mathcal{N}_{mp}= [4 - 4 e^{-2|\alpha|^2}]^{-1/2},\\
&\mathcal{N}_{mm}= [4 + 4 e^{-2|\alpha|^2} - 8 e^{-2|\alpha|^2}\cos(|\alpha|^{2})]^{-1/2}.
\end{aligned}
\label{equ:11}
\end{equation}
Here we consider the case without dissipation. 
A joint projective measurement of the two superconducting qubits in the $\sigma_x$
basis is then performed on the state $|\psi(t)\rangle$ 
(e.g., projecting onto $|+\rangle_{q_1}|+\rangle_{q_2}$ and $|-\rangle_{q_1}|-\rangle_{q_2}$), 
causing the magnon to collapse into the ideal four-components Shrödinger 
cat code state.
\par
This relative phase difference $\phi$ controls the direction of the 
displacement operations in phase space, thereby 
determining the symmetry of the generated cat states. 
The relative phases among the coherent components are preserved during the evolution, 
which guarantees robust quantum coherence and provides intrinsic protection against decoherence. 
Specifically, when $\phi = 0$, the state $|\psi(t)\rangle$ reduced to a 
two-component cat state. In contrast, for $\phi =  (2k+1)\pi/(4n_0-2)$ ($k \in \mathbb{Z}$), 
the four coherent states 
are distributed along orthogonal axes in phase space, resulting in the
$C_4$-symmetric 4C states.
This phase-controlled tunability provides a powerful means to engineer the 
superposition structure and phase-space symmetry of target 4C states 
in hybrid magnon-qubit systems.
\section{Wigner Functions and Quantum Master Equations}
The quantum interference and coherence effects in the generated cat states can 
be revealed by the rotated quadrature operator 
\cite{walls1994gj} and the 
Wigner function \cite{BUZEK19951}. 
The Wigner function is a phase-space quasiprobability 
distribution \cite{scully_zubairy_1997,LEONHARDT199589}.
The Wigner function for the magnon, described by the density matrix 
$\rho_{m}$, is defined as
\begin{equation}
  W ( \alpha)=\frac{2} {\pi} \langle \hat{\mathbf{D}} (\alpha) P \hat{\mathbf{D}} (-\alpha) \rangle, 
\end{equation}\label{equ:12}
where 
$\hat{\mathbf{D}} (\alpha)$
is 
the displacement operator; $P=e^{i \pi m^{\dagger} m}$ 
is the parity operator; and $\langle\cdots \rangle$ 
denotes the expectation value with respect to the quantum state \cite{BUZEK19951,PhysRevA.99.022302,walls1994gj}. 
We numerically simulate the Wigner functions of the magnon mode in the effective frame
at $t=40$ ns without the dissipation. 
As shown in Figs. \ref{fig2}(a)-(d), 
4C states with coherent amplitude $|\alpha|= 1.007$ can be obtained.
Adjacent coherent state components of the magnon mode are aligned along orthogonal axes, while 
opposite components are located at diametrically opposite points in phase space. 
These features are consistent with the 4C
states under the chosen parameters \cite{doi:10.1126/science.aaz9236,PhysRevLett.130.193603}.
\par
We further investigate the impact of system dissipation on the generation
of the 4C states. 
In a zero-temperature environment, the system-bath 
coupling leads to dissipative dynamics described by the Lindblad master equation:
\begin{equation}
  \begin{aligned}
    \frac{\partial\rho }{\partial t}=&-i\left[H_{\rm total},\rho\right]+\sum_{j} \frac{1}{2}\mathcal{L}_{c_j} \left[\rho\right],
  \end{aligned}\label{equ:13}
\end{equation}%%%%%%公式
where $\rho $ is the density matrix of the system in the original (laboratory) frame, 
and the standard Lindblad superoperator is defined as
$
\mathcal{L}_{o} [ \rho] =2o \rho o^{\dagger}-( o^{\dagger} o \rho+\rho o^{\dagger} o ) ;
$ 
the set of collapse operators is given by 
$c_j \in \{\sqrt{\kappa_m}m, \sqrt{\gamma_{q_1}}\sigma_1^{-},\sqrt{\gamma_{q_2}}\sigma_2^{-},\sqrt{\kappa_{a}}a\}$ , 
corresponding to the magnon, the two superconducting qubits, and the microwave cavity mode. 
The associated decay rates are
$\{\kappa_m, \gamma_{q_1},\gamma_{q_2},\kappa_{a}\}$ 
, representing the dissipation rates of each mode.
The dissipative dynamics governed by the effective Hamiltonian is obtained by 
deriving an effective master equation in the rotating frame.
By performing the transformation 
\begin{equation}
  U_{\mathrm{tot}}(t) = U_1(t) U_2(t) U_{\mathrm{aux}}^{\dagger}(t) e^{-S},
   \label{equ:14}
\end{equation} 
the fast oscillating terms are eliminated, 
while the slow dynamics that dominantly govern the system evolution are retained.
Therefore, 
the effective master equation can be written as
\begin{equation}
  \begin{aligned}
    \frac{\partial\tilde{\rho} }{\partial t}=&-i\left[H_{\rm eff},\tilde{\rho}\right]+\sum_{j} \frac{1}{2}\mathcal{L}_{a_j} \left[\tilde{\rho}\right],
  \end{aligned}\label{equ:15}
\end{equation}%%%%%%公式
where $\tilde{\rho}$ denotes the density matrix of the system in 
the transformed (effective) frame governed by the effective Hamiltonian $H_{\rm eff}$;  
$a_j$ are the collapse operators including qubit decoherence terms $\sqrt{\gamma_{q_j}/2} 
\sigma_j^x$, $\sqrt{\gamma_{q_j}/2} J_2(\mu_j) \sigma_j^y$, $\sqrt{[J_1^2(\mu_j) 
+ J_3^2(\mu_j)] \gamma_{q_j}/2} \sigma_j^z$; magnon damping $\sqrt{\kappa_m} m$; 
and a hybrid dissipation channel $\sqrt{\kappa_a} (G/\delta \sigma_1^z + G/\delta 
\sigma_2^z e^{-i\Phi} - g_3/\Delta_{\rm cm} m)$.  
The hybrid dissipation channel is obtained through adiabatic elimination 
of the cavity mode and is used to capture the correlated decay involving the qubits and magnon.
The detailed derivation can be found in the Appendix \hyperref[appendix:B]{B}.
\par
\begin{figure}%[htbp] %htbp 代表图片插入位置的设置
\centering %图片居中      
\includegraphics[width=8cm]{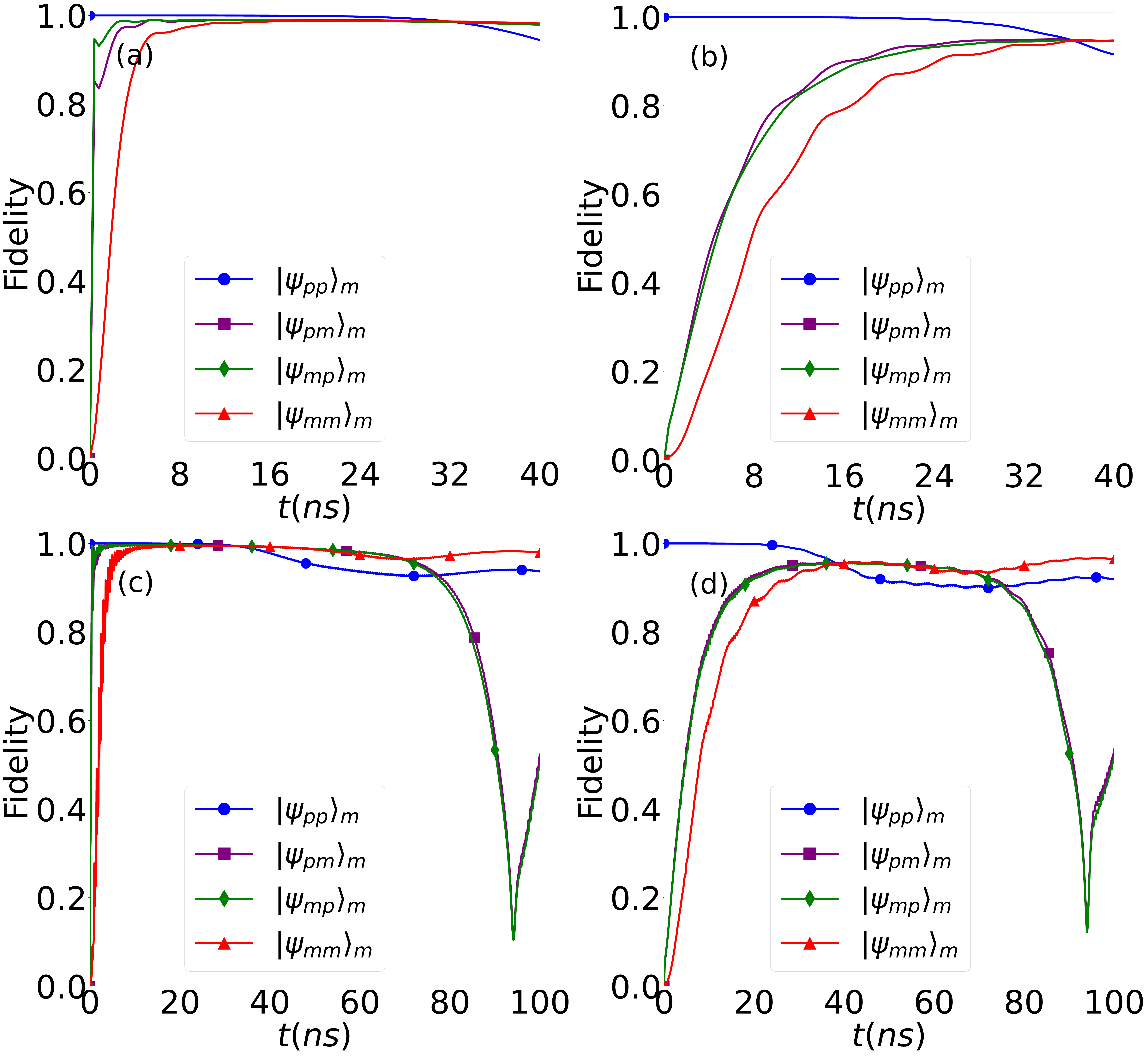} %[]中可选参数，可以设置图片的宽高      
%添加图体     
\caption{
The fidelity between two evolved states 
governed by the full Hamiltonian in Eq. (\ref{equ:1}) and the 
effective Hamiltonian in Eq. (\ref{equ:7}) is shown 
for different initial states of the magnon and cavity mode: 
small-amplitude coherent states in (a) and (c), and vacuum states in (b) and (d).
In Figs. (a) and (b), The parameters used  are matching those in Fig. \ref{fig2};
while the parameters used in (c) and (d) are alternative parameters: 
$\omega_f/2\pi = 8.0023\mathrm{GHz}$, 
$\Omega_f/\omega_f = 0.92$, $\omega_c/2\pi = 7.827\mathrm{GHz}$, 
$\omega_m/2\pi = 8\mathrm{GHz}$, $\omega_{q_1}=\omega_{q_2} = 0$, 
$g_1/2\pi = g_2/2\pi = 120\mathrm{MHz}$, $g_3/2\pi = 20\mathrm{MHz}$,$n_0=1$. 
\label{fig4}
}
\end{figure}
We utilized the QuTiP package \cite{JOHANSSON20121760,JOHANSSON20131234}
to study the system by numerical solution of the Eq. (\ref{equ:15}). 
In Fig. \ref{fig3}, each subplot shows the fidelity between the ideal 4C states 
$|\psi_{pp}\rangle_m$ or $|\psi_{mm}\rangle_m$ and the 
corresponding states evolved under various dissipation conditions.
The fidelity is plotted as a function of one dissipation rate, with the x-axis of each subplot 
corresponding to $\gamma_q$, $\kappa_m$, or $\kappa_c$, 
while the second dissipation parameter is varied across the different curves 
($0$ MHz, $0.25$ MHz, $0.5$ MHz, $0.75$ MHz, and $1.0$ MHz), and the third 
dissipation parameter is fixed at $0.5$ MHz.
Obviously, the states $|\psi_{pp}\rangle_m$ and $|\psi_{mm}\rangle_m$ exhibit similar sensitivity patterns.
Additionally, we observe that both components decrease as dissipation increases, 
particularly under simultaneous strong qubit and magnon decoherence. 
Whereas the impact of cavity loss $\kappa_c$ is relatively minor within the same range. 
This illustrated that the generation of 4C states is intrinsically robust 
against cavity-mode decay, the cavity plays a virtual role in mediating interactions. 
The robustness can be advantageous for actual implementations, 
where cavity modes often suffer from relatively fast energy loss.
\section{DISCUSSIONS}
In our work, we have set $n_0=1$. 
Next, we delve into the practicality of our scheme under those parameters
by analyzing the fidelity between the dynamics governed by the full 
Hamiltonian and those from the effective Hamiltonian.
\par
The time-independent anti-Hermitian generator 
$S$
used in the Nakajima transformation acts on the vacuum state and induces small 
displacements in the phase-space.
In the original frame, the initial states of the magnon $|\psi_{m,int}\rangle = U_{tot}(t)|0\rangle_m$ 
and cavity $|\psi_{a,int}\rangle = U_{tot}(t)|0\rangle_a$
are coherent states with the small-amplitudes.
These small displacements can theoretically remain negligible under the condition  
$g_3/\Delta_{\rm cm}\ll 1 $ and $G/\delta \ll 1$. 
We plot the time evolution of the fidelity in Fig. \ref{fig4} for 
initial small-amplitude coherent states 
[Figs. \ref{fig4}(a) and \ref{fig4}(c)] and initial vacuum states [Figs. \ref{fig4}(b) and \ref{fig4}(d)]. 
Comparing Fig. \ref{fig4}(a) with Fig. \ref{fig4}(b), and Fig. \ref{fig4}(c) with Fig. \ref{fig4}(d), 
we find that the initial slight phase-space displacement reducing fidelity 
but still yielding 4C states. 
The initial vacuum and the initial small-amplitude coherent states correspond 
to Gaussian wave packets centered at different points in phase space. 
Under a choosen relative phase $\phi$, the resulting dynamics preserve the 
symmetry of the coherent-state constellation; the two cases with different initial states
differ mainly by a global phase-space translation of the motional trajectory. 
The observed fidelity oscillations arise from the time-dependent coherent amplitude 
$\alpha(t)$ of the coherent components.
\par 
In Figs. \ref{fig4}(a) and \ref{fig4}(b), the parameters 
are set the 
same as those in Fig. \ref{fig2}. 
In Figs. \ref{fig4}(c) and \ref{fig4}(d), we investigate alternative parameter regimes.
The 4C states $|\psi_{pp}\rangle_m$ and $|\psi_{mm}\rangle_m$ can be successfully 
generated after a certain evolution time, 
while $|\psi_{pm}\rangle_m$ and $|\psi_{mp}\rangle_m$ emerge periodically. 
The drive parameters 
are set to $\omega_f/2\pi = 8.0023  \mathrm{GHz}$ and 
$\Omega_f / \omega_f = 0.92$. 
Additional system parameters include: 
cavity frequency $\omega_c / 2\pi = 7.827 \mathrm{GHz}$,
magnon frequency $\omega_m/ 2\pi = 8 \mathrm{GHz}$, 
qubit-cavity coupling strength $g_1/ 2\pi = g_2/ 2\pi = 120\mathrm{MHz}$, 
and magnon-cavity coupling strength $g_3/ 2\pi = 20 \mathrm{MHz}$. 
Numerical simulations show that within a wide frequency range of the magnon mode, it 
is possible to identify viable parameter sets that satisfy the requirements of our theoretical 
approximations. 
The frequencies of the superconducting qubit are set as $\omega_{q_j} = 0$ to simplify the effective 
dynamics in the rotating frame, as discussed in Appendix \hyperref[appendix:B]{B}.
\begin{figure}%[htbp] %htbp 代表图片插入位置的设置
\centering %图片居中      
\includegraphics[width=8cm]{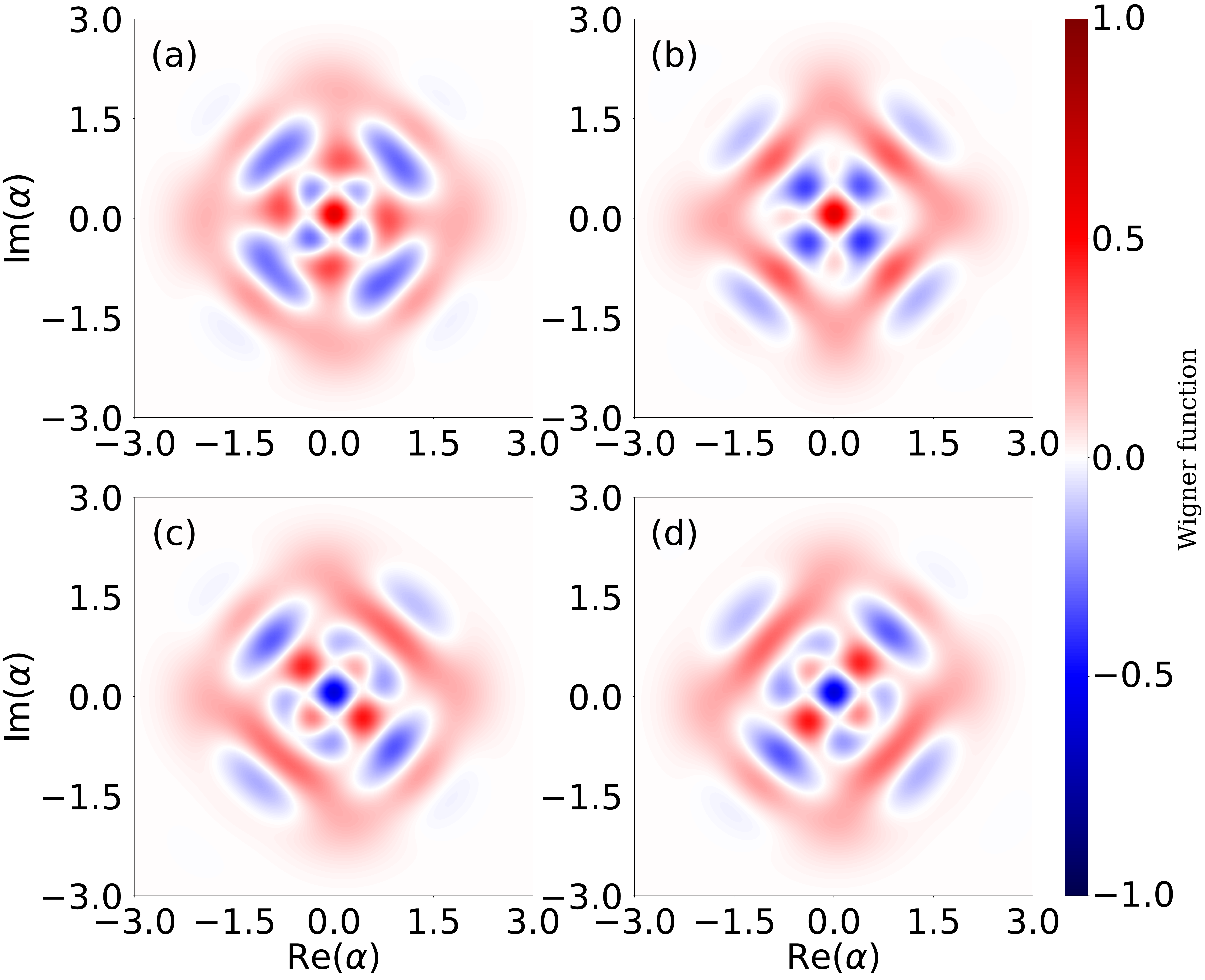} %[]中可选参数，可以设置图片的宽高      
%添加图体      
\caption{
Wigner functions of the 4C magnon states 
(a) $|\psi_{pp}\rangle_m$ 
(b) $|\psi_{mm}\rangle_m$ 
(c) $|\psi_{pm}\rangle_m$ 
(d) $|\psi_{mp}\rangle_m$ 
governed by the total Hamiltonian at $t = 50$ ns, where 
the initial states of 
the magnon mode and cavity mode are both vacuum. The coherent amplitude $|\alpha|$ is $1.2596$. 
The parameters are set the same as those in Figs. \ref{fig4}(c) and \ref{fig4}(d)
\label{fig5}
}
\end{figure}
\par
On the other hand, we plot the Wigner function of the 4C states in Fig. \ref{fig5} at 
$t=50$ ns with the initial vacuum states of the magnon and the microwave cavity mode in the original frame. 
Obviously, the 4C states can still be obtained. 
The parameters are matching those in the Figs. \ref{fig4}(c) and \ref{fig4}(d) with
coherent amplitude $|\alpha|=1.2596$. 
Our numerical simulations demonstrate the generating of 4C states by Floquet-engineering, 
while practical implementation in platforms such as spin-qubit-oscillator or
superconducting-qubit-oscillator architectures will require tailored drive amplitudes and 
detunings to optimize coherence and fidelity.
\section{CONCLUSION}
In summary, we have proposed a Floquet-engineering scheme 
to generate the 4C states in a 
hybrid ferromagnet-superconductor system.
Coherent magnon-qubit coupling is mediated via appropriately selected Floquet sidebands 
and the virtual-photon excitation of the cavity.
We analytically derived the master equation under the 
dissipation and verified the validity of the approximations.
Through the numerical 
simulations, 
we find that the generated 4C states exhibit strong robustness against cavity decay, 
as the cavity acts solely as a virtual mediator 
without participating in real energy exchange.
Moreover, the symmetry of the coherent-state constellation is controlled by the Floquet 
driving phase difference $\phi$, 
which protects the relative phases among the coherent components and thereby enhances coherence.
We also examined the impact of different initial states and system parameters.
These results highlight the potential of Floquet engineering in solid-state 
hybrid platforms for generating 4C nonclassical states and 
advancing robust quantum state manipulation with magnons. 
This work enables quantum information processing with magnon-based 
cat states in circuit QED, with applications in quantum error correction and simulation.
\par
\section*{ACKNOWLEDGEMENTS}
We thank Wenlin Li and Yexiong Zeng for the constructive discussions. 
This work was supported by the National Natural Science 
Foundation of China (Grant Nos. 12274053 and 62471089) and 
by the Fundamental Research 
Funds for the Central Universities (Grant Nos. 04442024074). 
\bibliography{main_text_file}

\onecolumngrid  % ← 切换为单栏
\appendix

\section*{Appendix A: Detailed Calculations of the Hamiltonian $H^{\prime}_{\rm fram}$}\label{appendix:A}
To derive the Hamiltonian $H^{\prime}_{\rm fram}$, we first apply the unitary 
transformation $U_2(t) = \exp(-i \omega_c a^\dagger a t - i \omega_m m^\dagger m t)$.
We then expand the
the time-dependent terms using the Jacobi-Anger expansion:
\begin{equation}
\begin{aligned}
\cos[2 \theta_1(t)] &= J_0\left(\mu_1\right) 
+ 2 \sum_{n = 1}^{\infty} J_{2n}\left(\mu_1\right) 
\cos\big(2n \omega_{f_1} t\big), \\[2pt]
\sin[2 \theta_1(t)] &= 2 \sum_{n = 1}^{\infty} J_{2n-1}\left(\mu_1\right) 
\sin\big[(2n{-}1) \omega_{f_1} t\big], \\[4pt]
\cos[2 \theta_2(t)] &= J_0\left(\mu_2\right) 
+ 2 \sum_{n = 1}^{\infty} J_{2n}\left(\mu_2\right) 
\cos\big[2n (\omega_{f_2} t + \phi)\big], \\[2pt]
\sin[2 \theta_2(t)] &= 2 \sum_{n = 1}^{\infty} J_{2n-1}\left(\mu_2\right) 
\sin\big[(2n{-}1) (\omega_{f_2} t + \phi)\big],
\end{aligned}
 \label{equ A1}\tag{A1}
\end{equation}
with $J_n(\mu)$ being the Bessel function of the first kind. In the rotating frame defined by the operator 
$U_{2}(t) = {\rm exp}(-i \omega_{c}a^{\dagger} a t- i\omega_{m}m^{\dagger}m t)$
The Hamiltonian $H_{\rm fram}$ in the interaction picture takes the form as
\begin{equation}
\begin{aligned}
H_{\text{fram}} = \; & 
\frac{\omega_{q_1}}{2} \left\{
    \sigma_1^z \left[ J_0\left( \mu_1 \right) + 2 \sum_{n=1}^\infty J_{2n}(\mu_1) \cos(2n\omega_{f_1}t) \right]
  + 2\sigma_1^y \sum_{n=1}^\infty J_{2n-1}(\mu_1) \sin[(2n-1)\omega_{f_1}t]
\right\} \\[0.5em]
& +
\frac{\omega_{q_2}}{2} \left\{
    \sigma_2^z \left[ J_0(\mu_2) + 2 \sum_{n=1}^\infty J_{2n}(\mu_2) \cos(2n(\omega_{f_2}t+\phi)) \right]
  + 2\sigma_2^y \sum_{n=1}^\infty J_{2n-1}(\mu_2) \sin[(2n-1)(\omega_{f_2}t+\phi)]
\right\} \\[0.8em]
& +
g_1 a e^{-i\omega_c t} \left\{
    \sigma_1^+ - \frac{i}{2}\sigma_1^y 
  - 2i\sigma_1^z \sum_{n=1}^\infty J_{2n-1}(\mu_1) \sin[(2n-1)\omega_{f_1}t]
  + \frac{i}{2}\sigma_1^y \left[ J_0(\mu_1) + 2 \sum_{n=1}^\infty J_{2n}(\mu_1) \cos(2n\omega_{f_1}t) \right]
\right\} \\[0.8em]
& +
g_1 a^\dagger e^{i\omega_c t} \left\{
    \sigma_1^+ + \frac{i}{2}\sigma_1^y 
  + i\sigma_1^z \sum_{n=1}^\infty J_{2n-1}(\mu_1) \sin[(2n-1)\omega_{f_1}t]
  - \frac{i}{2}\sigma_1^y \left[ J_0(\mu_1) + 2 \sum_{n=1}^\infty J_{2n}(\mu_1) \cos(2n\omega_{f_1}t) \right]
\right\} \\[0.8em]
& +
g_2 a e^{-i\omega_c t} \left\{
    \sigma_2^+ - \frac{i}{2}\sigma_2^y 
  - i\sigma_2^z \sum_{n=1}^\infty J_{2n-1}(\mu_2) \sin[(2n-1)(\omega_{f_2}t+\phi)]\right.\\
  &\left.+ \frac{i}{2}\sigma_2^y \left[ J_0(\mu_2) + 2 \sum_{n=1}^\infty J_{2n}(\mu_2) \cos(2n(\omega_{f_2}t+\phi)) \right]
\right\} \\[0.8em]
& +
g_2 a^\dagger e^{i\omega_c t} \left\{
    \sigma_2^+ + \frac{i}{2}\sigma_2^y 
  + i\sigma_2^z \sum_{n=1}^\infty J_{2n-1}(\mu_2) \sin[(2n-1)(\omega_{f_2}t+\phi)]\right.\\
  &\left.- \frac{i}{2}\sigma_2^y \left[ J_0(\mu_2) + 2 \sum_{n=1}^\infty J_{2n}(\mu_2) \cos(2n(\omega_{f_2}t+\phi)) \right]
\right\} \\[0.8em]
& +
g_3 \left( m a^\dagger e^{i\Delta_{\rm cm} t} + m^\dagger a e^{-i\Delta_{\rm cm} t} \right)
\end{aligned}
 \label{equ A2}\tag{A2}
\end{equation}
where $ \Delta_{\rm cm} = \omega_c - \omega_m$.
This Hamiltonian contains oscillating terms that differ by 
frequencies of $n\omega_{f_j}$, with $n$ being an integer.
We assume $\omega_c\gg g_{j},g_{j}J_0\left(\mu_j\right)/2,$ and choose a driving 
frequency $\omega_{f_j}$
that satisfies $\omega_{f_j} \gg \omega_{q_j}J_{2n}\left(\mu_{j}\right)/2, 
\omega_{q_j}J_{2n-1}\left(\mu_{j}\right)/2$ and $2n\omega_{f_j}-\omega_c \gg g_{j}J_{2n}(\mu_j) /2$, 
where $J_\ell(x)$ is the Bessel function of the first kind and $\ell$ is an integer. 
For a given drive frequency $\omega_{f_j}$, 
the parameter $n_0$ should be chosen such that
the corresponding $J_{2n_0-1}(\mu_j)$ term is the minimally detuned terms,
i.e., $2n_0-1 = Round[(\omega_c-\delta)/\omega_{f_j}]$.
This chosen term satisfies the large-detuning 
condition, $\delta \approx 5g_j J_{2n_0-1}(\mu_j)$, and dominates 
the qubit-cavity interaction in Eq. (\ref{equ:4}). 
Hereafter, we set 
$\omega_q = 0$ for simplicity. 
\section*{Appendix B: derivation of the effective master equation}\label{appendix:B}
First, the effective Hamiltonian and the full Hamiltonian satisfy the relation
$
H_{\mathrm{eff}} = U_{\mathrm{tot}}^{\dagger}(t) H_{\mathrm{tot}} U_{\mathrm{tot}}(t),
$
where the unitary transformation is shown in Eq. \ref{equ:14}
Here, $ U_1(t) $ and $ U_2(t) $ account for the Floquet modulation and interaction picture, 
respectively, while $ U_{\rm aux}(t) = \exp(-i H_0 t) $ and $ e^{-S} $ implement the 
rotating frame and perturbative transformation in Eq. (\ref{equ:5}).
To derive the effective master equation in Eq. (\ref{equ:15}), we transform the Lindblad 
master equation in Eq. (\ref{equ:13}) into the effective frame using the unitary operator $U_{\rm tot}(t)$:
\begin{equation}
\begin{aligned}
\frac{\partial \tilde{\rho}}{\partial t} = &\ \sum_{j=1}^2 \frac{\gamma_{q_j}}{2} \left\{ 2 \left[ \sigma_j^- + \frac{i \sigma_j^y}{2} + \frac{i \sigma_j^z}{2} \sin(2 \theta_j(t)) - \frac{i \sigma_j^y}{2} \cos(2 \theta_j(t)) \right] \tilde{\rho} \left[ \sigma_j^+ - \frac{i \sigma_j^y}{2} - \frac{i \sigma_j^z}{2} \sin(2 \theta_j(t)) + \frac{i \sigma_j^y}{2} \cos(2 \theta_j(t)) \right] \right. \\
&\ \left. - \left[ \sigma_j^+ - \frac{i \sigma_j^y}{2} - \frac{i \sigma_j^z}{2} \sin(2 \theta_j(t)) + \frac{i \sigma_j^y}{2} \cos(2 \theta_j(t)) \right] \left[ \sigma_j^- + \frac{i \sigma_j^y}{2} + \frac{i \sigma_j^z}{2} \sin(2 \theta_j(t)) - \frac{i \sigma_j^y}{2} \cos(2 \theta_j(t)) \right] \tilde{\rho} \right. \\
&\ \left. - \tilde{\rho} \left[ \sigma_j^+ - \frac{i \sigma_j^y}{2} - \frac{i \sigma_j^z}{2} \sin(2 \theta_j(t)) + \frac{i \sigma_j^y}{2} \cos(2 \theta_j(t)) \right] \left[ \sigma_j^- + \frac{i \sigma_j^y}{2} + \frac{i \sigma_j^z}{2} \sin(2 \theta_j(t)) - \frac{i \sigma_j^y}{2} \cos(2 \theta_j(t)) \right] \right\} \\
&\ + \frac{\kappa_c}{2} \left\{ 2 \left[ \frac{G}{\delta} \sigma_1^z + \frac{G}{\delta} \sigma_2^z e^{-i \Phi} - \frac{g_3}{\Delta_{\rm cm}} m \right] \tilde{\rho} \left[ \frac{G}{\delta} \sigma_1^z + \frac{G}{\delta} \sigma_2^z e^{i \Phi} - \frac{g_3}{\Delta_{\rm cm}} m^\dagger \right] \right. \\
&\ \left. - \left[ \frac{G}{\delta} \sigma_1^z + \frac{G}{\delta} \sigma_2^z e^{i \Phi} - \frac{g_3}{\Delta_{\rm cm}} m^\dagger \right] \left[ \frac{G}{\delta} \sigma_1^z + \frac{G}{\delta} \sigma_2^z e^{-i \Phi} - \frac{g_3}{\Delta_{\rm cm}} m \right] \tilde{\rho} \right. \\
&\ \left. - \tilde{\rho} \left[ \frac{G}{\delta} \sigma_1^z + \frac{G}{\delta} \sigma_2^z e^{i \Phi} - \frac{g_3}{\Delta_{\rm cm}} m^\dagger \right] \left[ \frac{G}{\delta} \sigma_1^z + \frac{G}{\delta} \sigma_2^z e^{-i \Phi} - \frac{g_3}{\Delta_{\rm cm}} m \right] \right\} - i [H_{\rm eff}, \tilde{\rho}].
\end{aligned}
\label{equ B2}\tag{B2}
\end{equation}
Here, we have $U_j^{\dagger}(t)\sigma_j^-U_j(t) = \sigma_j^- 
+ i\sigma_j^y/2 + i\sigma_j^z\sin(2\theta_j(t))/2-i\sigma_j^y\cos(2\theta_j(t))/2 $. 
As $\delta\gg G$ and $\Delta_{\rm cm}\gg g_3$, we retain terms up to first order in the perturbative 
expansion, yielding $e^{s}ae^{-s} = a + [S, a]$ and $e^{s}me^{-s} = m + [S, m]$. 
To evaluate terms like \(\sin[2 \theta_j(t)] \sin[2 \theta_k(t)]\), 
where \(\theta_j(t) = \frac{\Omega_{f_j}}{\omega_{f_j}} \sin(\omega_{f_j} t)\), we use 
the Jacobi-Anger expansion Eq. (\ref{equ A1}). 
Using Euler’s formula, the product $\sin[2 \theta_j(t)] \sin[2 \theta_k(t)]$ yields a 
double sum of terms $e^{i (n + m) \omega_{f_j} t}$. Under the RWA, we 
retain only non-oscillating terms ($n + m = 0$). 
On the other hand, based on our chosen driving frequency and intensity, we find that the Bessel function values are  
\( J_{2n-1} \approx 0.582, 0.1, 0.0047 \) for \( n = 1,2,3 \), respectively.  
Therefore, the leading contributions come from the first few terms (\( n \leq 2 \)), while higher-order harmonics are negligible.  
This hierarchy reflects the rapid decay of Bessel functions for increasing order when the argument is 
moderate, justifying the truncation. 
After applying the RWA and truncating higher-order terms, the effective master equation in Eq. (\ref{equ:15}) 
is obtained, with collapse operators as in the main text. 
\end{document}